\documentclass[prl,twocolumn]{revtex4-1}

\usepackage {amssymb}
\usepackage{graphicx}% Include figure files
\usepackage {color}

\newcommand{\be}{\begin{eqnarray}}
\newcommand{\ee}{\end{eqnarray}}
\newcommand{\bee}{\begin{eqnarray*}}
\newcommand{\eee}{\end{eqnarray*}}

\newcommand{\Z}{{\mathbb Z}}

\newcommand{\Name}[1]{\rm {#1}, }
\newcommand{\REVIEW}[4]{{\it {#1}}, {\bf {#2}} ({#3}) {#4}}

\begin{document}

\title []{Bloch oscillations and accelerated Bose-Einstein condensates in an optical lattice}

\author {Andrea SACCHETTI}

\address {Department of Physics, Informatics and Mathematics, University of Modena e Reggio Emilia, 
Modena, Italy\\Via G. Campi 213/A, Modena - 41125 - Italy}

\email {andrea.sacchetti@unimore.it}

\date {\today}

\begin {abstract} 
{We discuss the method for the measurement of the gravity acceleration $g$ by 
means of Bloch oscillations of an accelerated BEC in an optical lattice. \ This method has a theoretical critical point due to the fact that the period of 
the Bloch oscillations depends, in principle, on the initial shape of the BEC wavepacket. \ Here, by making use 
of the nearest-neighbor model for the numerical analysis of the BEC wavefunction, we show that in real experiments 
the period of the Bloch oscillations does not really depend on the shape of the initial wavepacket and that the relative 
uncertainty, due to the fact that the initial shape of the wavepacket may be asymmetrical, is smaller than the one 
due to experimental errors. \ Furthermore, we also show that the relation between the oscillation period and the 
scattering length of the BEC's atoms is linear; this fact suggest us a new experimental procedure for the 
measurement of the scattering length of atoms.}

\end{abstract}

\maketitle

%\normalsize
%
%{\bf Highlights:}
%
%\begin {itemize}
% \item [$\bullet$] Discrete nonlinear Sch\"odinger model for accelerated ultracold condensates in optical lattices;
%
%\item [$\bullet$] Numerical computation of wavefunction BECs dynamics for times of order of the Bloch oscillation period;
% 
% \item [$\bullet$] Correlation between oscillation period of the BEC's center of mass and nonlinearity strength;
%
%\item [$\bullet$] Discussione of the validity of the Bloch Theorem for accelerated BECs in an optical lattice
% 
%\end {itemize}
%
%{\bf Keywords:}
%
%\begin {itemize}
%
%\item [$\bullet$] Bose-Einstein condensates in periodic potentials
%
%\item [$\bullet$] Discrete nonlinear Schr\"odinger equations;
%
%\item [$\bullet$] Bloch Oscillations.
%
%\end {itemize}

%\pacs{03.75.Lm}{Tunneling, Josephson effect, Bose-Einstein condensates in periodic potentials, solitons, vortices, and topological excitations}
%\pacs{37.10.Jk}{Atoms in optical lattices}
%\pacs{02.70.Hm}{Spectral methods}

The idea of using accelerated ultracold atoms moving in an optical lattice \cite {Bloch1,RSN,SPSSKP,Shin} 
has opened the field to multiple applications. \ For instance, a cloud of accelerated ultracold bosons prepared in an essentially one-dimensional optical lattice exhibits 
Bloch oscillations \cite {DPRCS}; by making use of this property Clad\'e 
{\it et al} \cite {CGSNJB} (see also \cite {CladeReview,TinoReview} for a recent review) proposed a method for the measurement of the value for the 
gravity acceleration $g$ using ultracold bosons confined in a vertical optical 
lattice. \ More recently, Meinert {\it et al} \cite {MMKLWGN} have observed, for large values of the strength of the uniform acceleration, 
interaction-induced coherent decay and revival of the matter-wave quantum phase of the Bloch oscillating ensemble; when the strength of the uniform acceleration is reduced 
a transition from regular to quantum chaotic dynamics is observed. \ In that paper is also founded that the revival period
is entirely determined by the interaction strength and thus  provides a precise measure for the on-site interaction energy, and consequently for the scattering 
length. \ We 
also mention recent results where ultacold atoms have been used for the direct measurement of the universal Newton gravitation constant $G$ \cite {RSCPT} and 
of the gravity-field curvature \cite {RCSMPT}.

We focus now our attention on the basic idea proposed by Clad\'e {\it et al} \cite {CGSNJB}: when a one-dimensional Bose-Einstein condensate (BEC) 
in the vertical optical lattice performs periodic Bloch oscillations then the determination of the 
gravity acceleration $g$ can be obtained by measuring the period $T$ of the Bloch oscillations. \ Indeed, 
recalling from a seminal paper by Felix Bloch \cite {FB} (see also the textbook \cite {Callaway}) that for a single accelerated particle 
in an optical lattice the Bloch oscillation has period 
\begin{eqnarray}
T = \frac {2 \pi \hbar}{m g b} \, , \label {Eq0}
\end{eqnarray}
where $m$ is the mass of the particle, $\hbar$ is the Planck constant and $b$ is the lattice period, then a 
precise value of the gravity acceleration $g= \frac {2 \pi \hbar}{m b T}$ is obtained by means of the experimental measurement 
of the period $T$. \ The value for the gravity acceleration $g$ obtained by \cite {FPST,PWTAPT} was 
consistent with the one obtained by classical gravimeters; and the experimental result was affected by a relative 
uncertainty of order $6 \times 10^{-6}$, where the error's sources were mainly assumed to be a consequence of the 
experimental settings, e.g. the lattice's laser is not frequency stabilized, there is a small deviation from the 
vertical direction of the lattice, and so on. 

However, this ingenious method has, in principle, a theoretical weakness because the Bloch Theorem does not have a 
counterpart for BECs; indeed, Bloch oscillations are theoretically predicted only for a single accelerated particle 
in a periodic potential. \ Indeed, using gauge transformation applied to the time dependent Schr\"odinger equation for a 
single particle the tilt can be viewed as periodic driving of the system with Bloch frequency. \ In other words if $\psi (x,t) =
e^{i \lambda t/\hbar} \psi (x)$ is a solution of the time dependent Schr\"odinger equation for a 
single particle in a periodic field with a Stark perturbation then 
$ \psi_j (x,t) = e^{i (\lambda -j mg b) t/\hbar} \psi (x+jb)$, $j \in \Z$, is 
still a solution too and Stark-Wannier ladders occur \cite {WS}; then we expect that their linear combination is, up to a phase term, a 
periodic function with period $T$ given 
by (\ref {Eq0}). \ We must underline that this argument does not apply to the nonlinear Schr\"odinger equation, which describes the BEC's 
wavefunction; indeed, even in such a case we have families of solutions of the form $ \psi_j (x,t)$, for some $\psi (x)$, and 
thus a kind of Wannier-Stark ladders picture occurs, but we have to care of the fact that a linear combination of solutions of a given 
nonlinear equation is not, in general, a solution of such an equation. \ In fact, we 
expect that the nonlinear perturbation affects the oscillation period; this phenomenon can be seen in a simple double-well model with a 
nonlinear perturbation \cite {Jona}. \ In conclusion, in the case of accelerated BECs in an optical lattice oscillations are still 
expected with a given period still denoted by $T$, but the estimate of the effect of the atomic binary interactions on the 
oscillation period $T$ of the BEC must be considered \cite {S3,S4,KKG,WWMK}.  \ This fact opens a theoretical 
question concerning the validity of the method proposed for the determination of the gravity acceleration by means 
of the measurement of the oscillation period, as pointed out by \cite {S4} in a simple model where the 
potential lattice has a small number of wells. \ In fact, in order to minimize the effect of atomic binary 
interactions on the value of the period $T$, has been used, in the experiments above \cite {FPST,PWTAPT}, 
a BEC of a particular Strontium's isotope ${}^{88}Sr$ with a small scattering length $a_s$, and the initial 
wavepacket is prepared on a large number of wells, typically more than one hundred.

In this paper we numerically prove that the effect of the atomic binary interactions on the oscillation period 
is negligible when the BEC is initially prepared on a sufficiently large number of periodic 
cells. \ The nonlinear Sch\"odinger equation which describes the dynamics of the BEC wavefunction can be 
reduced, in the semiclassical limit, to a discrete nonlinear Schr\"odinger equation. \ The numerical solution to 
such a latter equation shows an oscillating behavior and the oscillation period can be numerically computed. \ We 
apply such a method where we choose the values of the physical parameters as in the experiment 
\cite {PWTAPT} and, in such a case, we find a relative uncertainty of the oscillation period of order 
$1 \times 10^{-6}$ due to the fact that the initial wavepacket may have an asymmetrical shape. \ Therefore we can 
conclude that the method proposed by Clad\'e {\it et al} \cite {CGSNJB} properly works for a vertical BEC initially 
prepared on a number of wells large enough. \ Furthermore, we also numerically show that the oscillation period 
linearly depends on the scattering length of the BEC's atoms; this result suggests a new experimental method for 
the measurement of the scattering length. \ In fact, the use of accelerated BECs in an optical lattice may be eventually 
used to measure the gravity acceleration $g$, as Clad\'e {\it et al} proposed, and even to measure the scattering 
length of the BEC's atoms instead of standard methods based on spectroscopy or, more recently, on two-component BECs 
\cite {Egorov}.

In order to discuss the dynamics of a one-dimensional cloud of cold bosons in a periodical optical 
lattice under the effect of the gravitational force we assume that the periodic potential has the usual shape
\begin{eqnarray}
V (x) = V_0 \sin^2 (k_L x) \label {Eq1}
\end{eqnarray}
where $b=\frac 12 \lambda_L$ is the period, $\lambda_L =\frac {2\pi }{k_L}$ and $V_0 = \Lambda_0 E_R$, where $E_R$ 
is the photon recoil energy. \ The one-dimensional BEC is governed by the one-dimensional time-dependent 
Gross-Pitaevskii equation with a periodic potential and a Stark potential
\begin{eqnarray}
i \hbar {\partial_t \psi} = H_B \psi +  m g x \psi + \gamma |\psi |^{2 } \psi \,  ,  \label {Eq2}
\end{eqnarray}
where the wavefunction $\psi (\cdot , t ) \in L^2 ({\mathbb R} ,dx)$ is normalized to one $\| \psi (\cdot , t ) \|_{L^2} 
= \| \psi_0 (\cdot )\|_{L^2} =1 $  and where 
\begin{eqnarray*}
H_B = - \frac {\hbar^2}{2m} \partial^2_{xx} + V (x) 
\end{eqnarray*}
is the Bloch operator with periodic potential $V (x)$; $\psi_0 (x) = \psi (x,0)$ is the initial wavefunction of 
the BEC. \ By $\gamma $ we denote  the one-dimensional nonlinearity strength. \ The study of the dynamics of the 
wavefunction $\psi$, solution to equation (\ref {Eq2}), is then obtained by means of a discrete nonlinear 
Schr\"odinger equation (DNLS). \ The idea is basically simple \cite {S4} and it consists in assuming that the 
wavefunction $\psi$, when restricted to the first band of the periodic Schr\"odinger problem, may be written as a 
superposition of vectors $u_\ell (x) = W_1 (x-x_\ell) $ localized on the $\ell -$the cell of the lattice, where 
$W_1$ is the Wannier function associated to the first band and $x_\ell = \ell \frac {\pi}{k_L}$ is the 
coordinate of the center of the $\ell -$th cell; that is 
\begin{eqnarray}
\psi (x,t) = \sum_{\ell \in {\mathbb Z}} c_\ell (t) u_\ell (x) \, . \label {Eq3Bis}
\end{eqnarray}
By means of such an approach the unknown functions $c_\ell (t)$ turn out to be the solutions to discrete nonlinear 
Schr\"odinger equations (DNLS) which dominant terms are given by (here we denote $\dot {} = \frac {d}{dt}$)
\begin{eqnarray}
&& i \hbar \dot c_\ell = - \lambda_D c_\ell - \beta \left ( c_{\ell +1} + c_{\ell -1} \right ) + \nonumber \\ 
&& \ \ + \gamma \| u_0 \|^{4}_{L^4} |c_\ell|^{2} c_\ell + 
m g  b \ell c_\ell \, , \ \ell \in {\mathbb Z} \, , \label {Eq4}
\end{eqnarray}
where $\lambda_D$ is the ground state of a single cell potential and where $\beta $ is the hopping matrix element 
between neighboring sites. \ In fact, in the semiclassical limit it turns out that $\lambda_D$ has dominant value 
given by $\hbar k_L \sqrt {V_0/2m}$, and the parameter $\beta$ is expected to be such that $4 \beta$ is equal to 
the amplitude of the first band. 

The theoretical question about the validity of the nearest-neighbor model (\ref {Eq3Bis})-(\ref {Eq4}) has been 
largely debated. \ A positive, and fully rigorous, answer to this question has been given in the semiclassical 
limit for BECs in a periodic potential \cite {FS,P1,P2} and, more recently, for accelerated BECs in a  
potential with a finite number of wells \cite {S4}. \ In fact, in our problem the semiclassical limit has to be 
understood as $\Lambda_0$ large enough and $\hbar $ fixed. \ A numerical evidence of the validity of the 
nearest-neighbor model consists in showing that the hopping matrix element between the sites of the lattice rapidly 
decreases when the number of cells between the sites increases. \ In particular, numerical experiments 
\cite {AKKS,EHLZCMA} suggest that the nearest-neighbor model properly works when $\Lambda_0 \ge 10$.

In typical experiments \cite {FPST,PWTAPT} the lattice period is $b=\lambda_L/2 = 266 \, nm$, that is $\lambda_L = 532 \, nm$; then, the Bloch period (\ref {Eq0}) 
predicted by the Bloch theorem is given by
\begin{eqnarray*}
T = \frac {2\pi \hbar}{m g b } = 1.740 \, ms \, .
\end{eqnarray*}
The lattice potential depth is $V_0 = \Lambda_0 \cdot E_R$, where $E_R = \frac {2 \pi^2 \hbar^2}{m\lambda_L^2} = 
50.38\, kHz \cdot \hbar$ is the photon recoil energy and $\Lambda_0$ is assumed to be large enough, typically 
$\Lambda_0 = 10$ such that the nearest-neighbor model (\ref {Eq3Bis})-(\ref {Eq4}) properly works. \ For such 
values it turns out that the first band of the Bloch operator $H_B$ has endpoints $ {E}_1^b = 4.32 \cdot E_R$ and 
$ {E}_1^t = 4.58 \cdot E_R$. \ Hence, the values of the width of the first band is given by $ B_1 := {E}_1^t -  
{E}_1^b = 0.26 \cdot E_R $ and $\beta = \frac 14 B_1 = 0.065 \cdot E_R $. \ In order to estimate the 
one-dimensional nonlinearity strength $\gamma$ it is expected that it is of the order \cite {SPR}
 \begin{eqnarray*}
 \gamma = \frac {\gamma_{3D}}{2 \pi d_\perp^2} 
 \end{eqnarray*}
where $d_\perp$ is the length of the transverse confinement, $\gamma_{3D} =\frac {4{\mathcal N} \pi a_s \hbar^2}{m}$ 
is the effective nonlinearity strength for the three-dimensional Gross-Pitaevskii equation, $a_s$ denotes the 
scattering length of the Strontium isotope ${}^{88}Sr$ and ${\mathcal N}$ is the number of atoms of the 
condensate. \ In typical experiments $d_\perp \approx 180 \cdot 10^{-6}\, m$ and ${\mathcal N} = 10^5 \div 
10^6$. \ Recently, it has been estimated that the ground-state s-wave scattering length $a_s$ for the main bosonic 
isotope of Strontium ${}^{88}Sr$ is such that $a_s = -a_0 \div 13 a_0$ \cite {MMSNCKPC}, where $a_0$ is the Bohr 
radius; more recent results suggest that $a_s = -2 a_0$ \cite {MMYDK}.

By introducing the slow time $\tau = \beta t /\hbar $ and by setting $d_\ell (\tau ) = c_\ell (t) 
e^{i\lambda_D t /\hbar }$ then the DNLS equation (\ref {Eq4}) takes the form (here we denote $' = \frac {d}{d\tau }$)
\begin{eqnarray}
i d_\ell' =  - (d_{\ell+1} + d_{\ell-1}) + \eta |d_\ell |^2 d_\ell +  \ell \rho d_\ell \, , \ 
\ell \in {\mathbb Z} \, , \label {Eq5}
\end{eqnarray}
where, when the values of the physical parameters are chosen as above, the adimensional parameters $\eta$ 
and $\rho$ take values
\begin{eqnarray}
\eta = \frac {\gamma \| W_1 \|_{L^4}^4}{\beta } = \frac {2{\pi} \Lambda_0^{1/4}   \hbar^2}{  b \beta d_\perp^2} 
\frac {{\mathcal N} a_s}{ m } 
= -0.03 \label {Eq6}
% -0.151 \cdot 10^{-1} \div 0.197
\end{eqnarray}
for ${\mathcal N}= 10^6$ and $a_s = - 2 a_0$, and  
\begin{eqnarray*} 
\rho = \frac {f b}{\beta} = \frac {m g b}{\beta} = 1.103 \, .
\end{eqnarray*}
In these units and for such a value of $\rho$ then the Bloch period (\ref {Eq0}), still denoted by $T$, is given by
\begin{eqnarray*}
T = \frac {2\pi \beta}{m g b} = \frac {2\pi}{\rho} = 5.696(45)
\end{eqnarray*}
in adimensional units.

\begin{center}
\begin{figure}
\includegraphics[height=8cm,width=8cm]{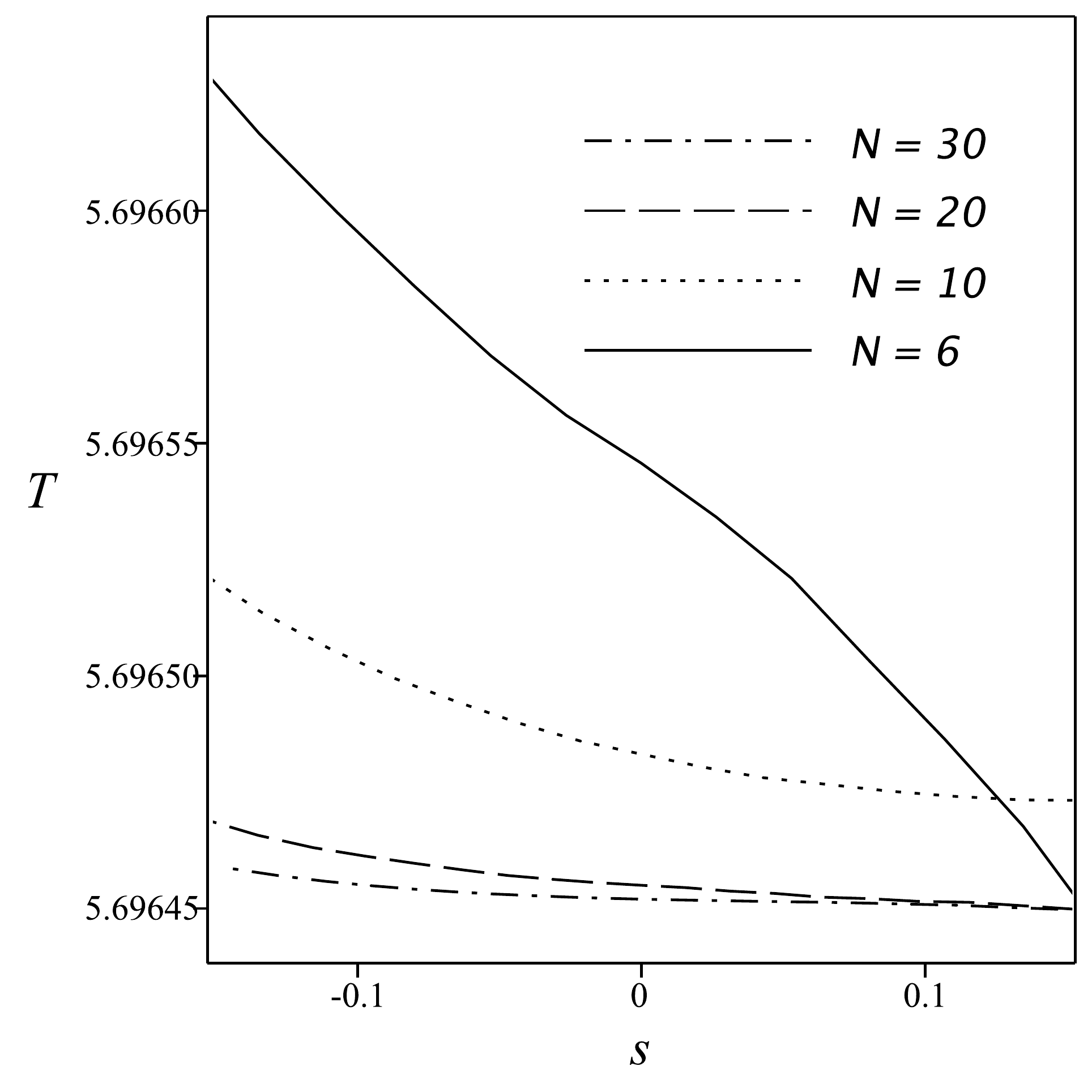}
\caption{\label {Fig2} In this figure we plot, for given values of $N$, the graphs of the period $T$ versus the adimensional parameter $s$, which 
measures the skewness of the initial wavepacket.}
\end{figure}
\end{center}

\begin{center}
\begin{figure}
\includegraphics[height=8cm,width=8cm]{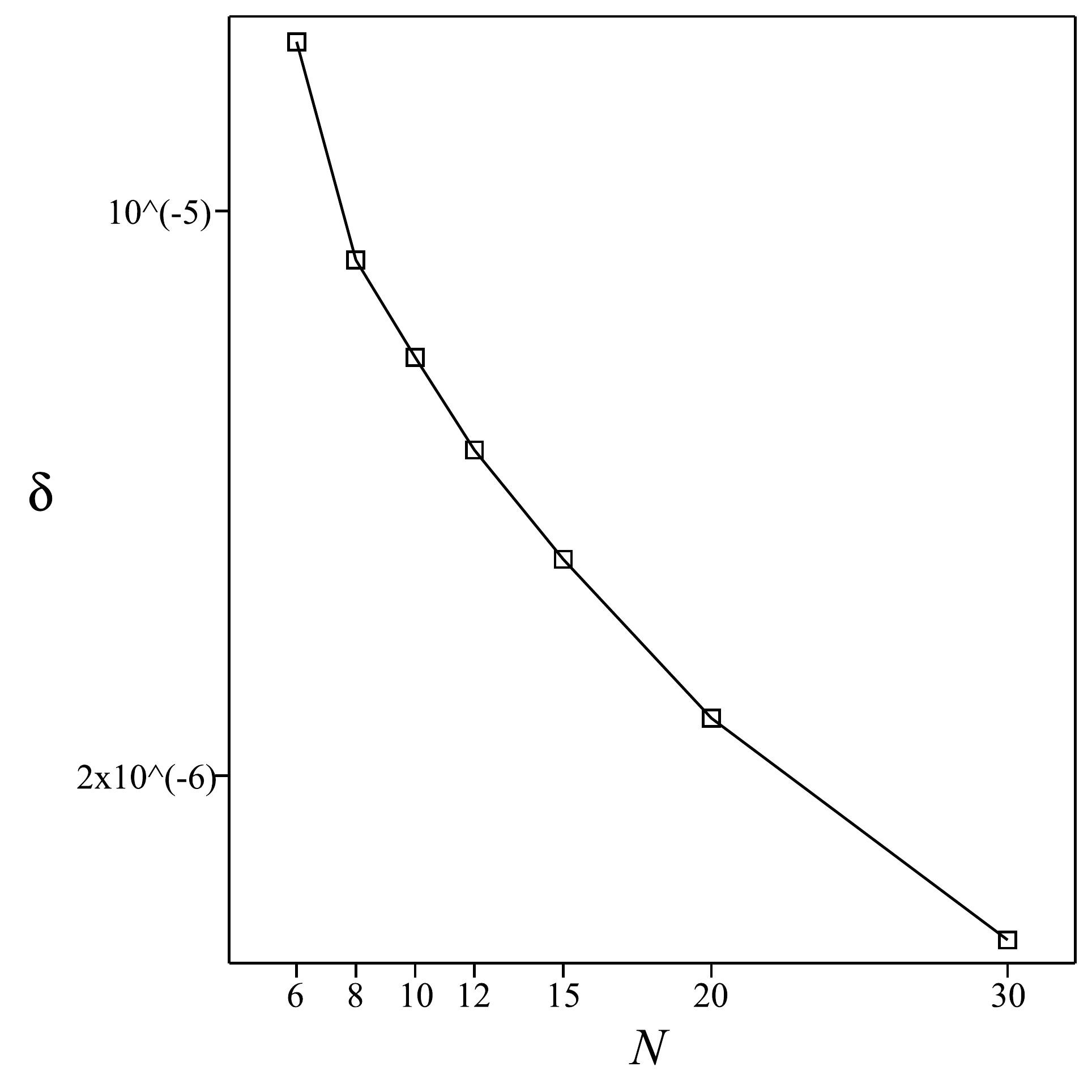}
\caption{\label {Fig3} Logarithmic plot of the relative uncertainty of the period for different number of wells 
when the adimensional nonlinearity strength (\ref {Eq6}) is fixed to the theoretical 
value $-0.03$ and when the skewness $s$ of the initial wavepacket runs in the interval $[-0.15,0.15]$.}
\end{figure}
\end{center}

We assume that the initial wavefunction $\psi_0 (x) = \psi (x,0)$ is initially prepared on a finite number $N$ 
of wells 
\begin{eqnarray*}
\psi_0 (x) = \sum_{\ell =0}^N c_\ell^0 u_\ell (x) 
\end{eqnarray*}
and we numerically compute the solutions to equation (\ref {Eq5}) with initial condition $c_\ell (0) = 
c_\ell^0$. \ The expectation value of the centroid of the wavepacket
\begin{eqnarray*}
\langle x \rangle^t = \langle \psi (x,t) | x \psi (x,t ) \rangle 
\end{eqnarray*}
exhibits an oscillating behavior with period $T $ depending, in principle, by the adimensional effective 
nonlinearity strength $\eta$, by the adimensional strength $\rho$ of the Stark potential and by the shape of the 
initial wavepacket $\psi_0$. \ However, here we show that when the initially wavepacket $\psi_0$ is initially 
prepared on a number $N$ of wells large enough (typically $N \ge 30$) then the period $T$ practically does 
not depend on these parameters and then the relative uncertainty is quite small. \ To this end we assume that the 
values $c_\ell^0$ are given by means of a binomial distribution with parameters $N$ and $p\in (0,1)$; that is
\begin{eqnarray*}
c_\ell^0 = C  {N \choose \ell} p^\ell (1-p)^{N-\ell }
\end{eqnarray*}
where $C$ is a normalization constant in order to have $\| \psi_0 \|=1$. \ Then we fix the strength $\eta$ to the 
theoretical value (\ref {Eq6}), corresponding to ${\mathcal N} =10^6$ atoms and scattering length $a_s = - 2 a_0$, 
and we vary the number $N$ of wells where the BEC wavefunction is initially prepared, and the skewness 
\begin{eqnarray*}
s = \frac {\langle \psi_0 | x^3 \psi_0 \rangle - 3 \mu \sigma^2 - \mu^3}{\sigma^3}
\end{eqnarray*}
of the initial wavepacket, which measures the asymmetry of the initial wavepacket (as usual $\mu = \langle 
\psi | x \psi_0 \rangle $ and $\sigma^2 = \langle \psi | x^2 \psi_0 \rangle - \mu^2$), corresponding to 
different values of $p$. \ In Figure \ref {Fig2} we plot the dependence of the period $T$ by the skewness 
parameter $s$ for $\eta$ fixed to the theoretical value $-0.03$ and for different values of $N$. \ It turns out 
that when $N$ is large enough then
\begin{eqnarray*}
\delta = \max_{s \in [-0.15,+0.15]} \frac {\Delta T}{T} = 1.25 \cdot 10^{-6} \, , \mbox { for } \ N =30.
\end{eqnarray*}
In particular, in Table \ref {Tab1} we collect the value of the relative uncertainty of the period for different 
values of $N$ and we plot these values in a logarithmic graphic in Figure \ref {Fig3}. \ It turns out that the 
relative uncertainty, due to the fact that the oscillation period depends on the initial shape of the wavepacket, 
rapidly decreases when the number $N$ of wells, where the initial wavepacket is prepared, increases; in particular, 
$\delta$ becomes much smaller that the relative uncertainty estimated by \cite {PWTAPT}, due to the experimental 
settings, when $N$ is larger or equal to $30$. 
\begin{table}
\begin{center}
\begin{tabular}{|c||c|} 
\hline
$N$ &  $\delta $ \\ \hline \hline
6 & $1.62 \cdot 10^{-5}$  \\ \hline
8 &  $8.69 \cdot 10^{-6}$  \\ \hline
10 & $6.59 \cdot 10^{-6}$ \\ \hline
12 & $5.06 \cdot 10^{-6}$ \\ \hline
15 & $3.71 \cdot 10^{-6}$ \\ \hline
20 & $2.36 \cdot 10^{-6}$ \\ \hline
30 & $1.25 \cdot 10^{-6}$ \\ \hline
\end{tabular}
\caption{Relative uncertainty of the period for different values of $N$ when the adimensional nonlinearity 
strength (\ref {Eq6}) is fixed to the theoretical value $-0.03$ and when the skewness $s$ of the initial wavepacket 
runs in the interval $[-0.15,0.15]$.}
\label{Tab1}
\end{center}
\end {table}
The physical qualitative intuition beyond such a result may have an explanation, at least when the scattering length $a_s$ is not negative; indeed, let the 
total number ${\mathcal N}$ of atoms of the BEC be fixed, as well as the scattering length $a_s$, then if the BEC normalized initial wavefunction $\psi_0$ is 
distributed on a sufficiently large number $N$ of cells then the repulsive binary interaction, associated to the nonlinear term  $\gamma |\psi_0 |^2 \psi_0$ of 
the Gross-Pitaevskii equation (\ref {Eq2}), can be assumed to be a rather small perturbative term at the initial instant and then it does not really affect the dynamics 
of the wavefunction $\psi (x,t)$ for times not too large. \ Hence, Bloch oscillations with period (\ref {Eq0}) would be observed. \ In the case of attractive binary interaction corresponding to 
$a_s <0$ then a similar argument would work provided that blow-up phenomena don't occur; in fact, functional estimates \cite {Car} prevent blow-up for such a 
model.

\begin{center}
\begin{figure}
\includegraphics[height=8cm,width=8cm]{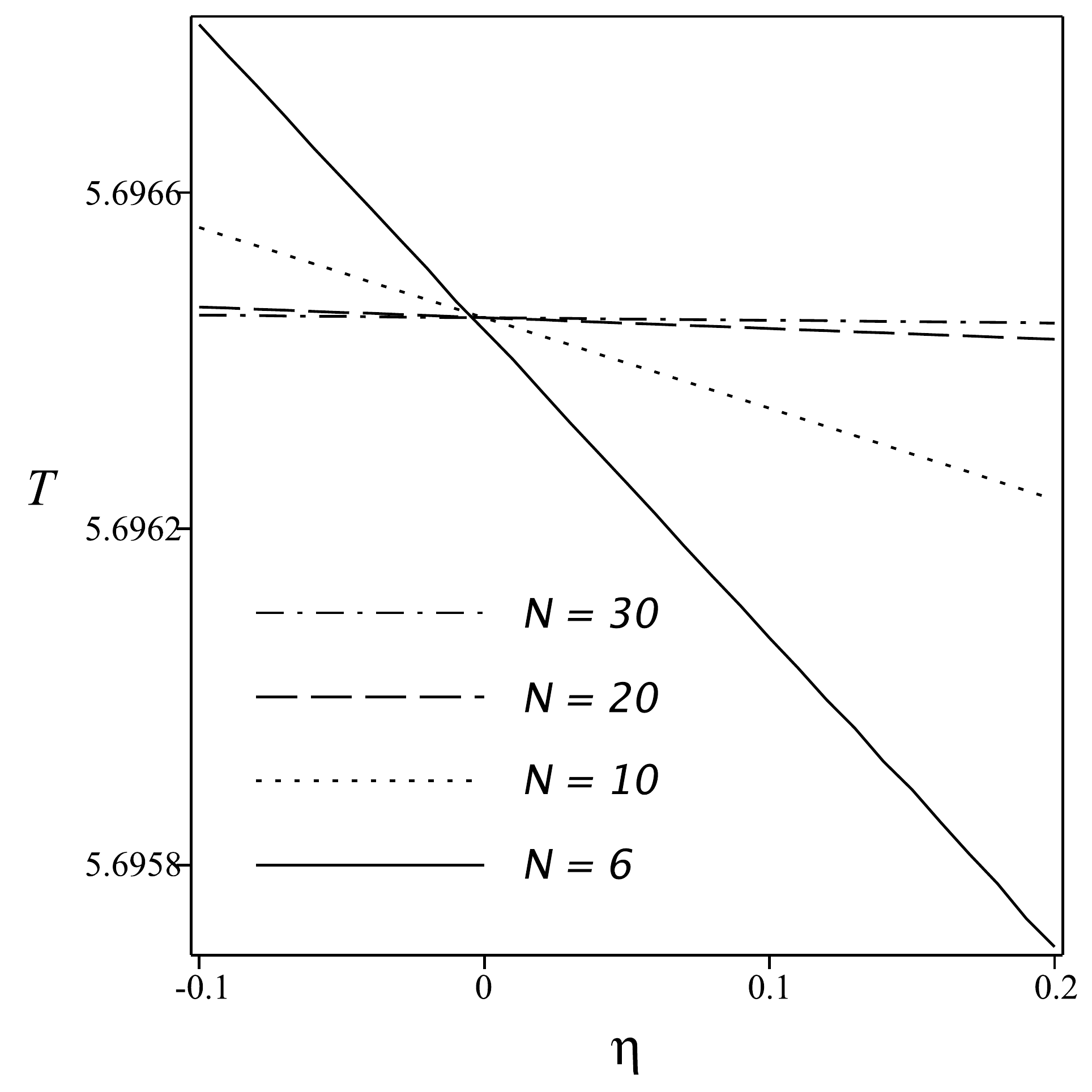}
\caption{\label {Fig1} In this figure we plot the 
graphs of the period $T$ versus the adimensional parameter $\eta$ for given values of $N$. \ In particular, it 
turns out that $T$ linearly depends by $\eta$. \ The slope of the straight line depends on $N$, in particular the 
values of the slope, for different values on the number $N$, are equal to $-3.66 \cdot 10^{-3}$ (for $N=6$), 
$-1.08 \cdot 10^{-3}$ (for $N=10$), $-1.28 \cdot 10^{-4}$ (for $N=20$) and $-3.09  \cdot 10^{-5}$ (for $N=30$).}
\end{figure}
\end{center}

Then we discuss how the oscillation period depends by the adimensional effective nonlinearity strength $\eta$. \ 
In Figure \ref {Fig1} we plot the dependence of the period $T$ by the adimensional nonlinearity strength $\eta$ in 
the interval $[-0.1,0.2]$ for different values of $N$ and when the initial wavepacket is 
symmetric, that is $p= \frac 12$. \ The adimensional strength $\rho$ of the Stark potential is fixed and equal to 
$1.103$. \ It turns out that the oscillation period \emph {linearly} depends on $\eta$. \ This fact suggests us a new 
experimental method for the measurement of the scattering length for bosonic 
atoms (not only for the isotope ${}^{88}Sr$). \ Indeed, by equation (\ref {Eq6}) and by the fact that $T$ 
linearly depends by $\eta$, then we can conjecture that the oscillation period depends on the scattering length 
$a_s$ and the number ${\mathcal N}$ of atoms as follows
\begin{eqnarray}
T = \frac {2\pi \hbar}{m g b} + c \frac {a_s {\mathcal N}}{m} \label {Eq8}
\end{eqnarray}
for some unknown constant $c$ which does not depend on the particular atom chosen, but $c$ only depends on 
the lattice potential, the transverse confinement, and the initial vertical confining potential, from which depends the initial shape 
of the wavefunction as well as the number $N$ of wells where the wavefunction is initially prepared; and where the value of $g$ is assumed now to be already known 
(by, e.g., a classical gravimeter). \ The value of $c$ can be obtained by measuring the oscillation period for an accelerated BEC 
in an optical lattice where the atoms of the BEC have a scattering length $a_s$ already known. \ Once 
we have experimentally obtained the value of $c$ then the experimental apparatus is ready to get the scattering 
length of any other atom by measuring the period when the BEC oscillates in a vertical optical lattice. 

In conclusion, we can state that the method proposed by Clad\'e {\it et al} \cite {CGSNJB} properly works  
when the initial wave packet is prepared on a sufficiently large number of wells; if not the period of the Bloch 
oscillations may be strongly affected by the strength on the nonlinearity term as well as by the 
skewness of the initial wavepacket. \ However, in the experiment realized by \cite {FPST,PWTAPT} the wavepacket is 
initially prepared on more than one-hundred wells and then we can conclude that the relative uncertainty, due to 
the fact that the oscillation period may depends on the shape of the initial wave-packet, is much less than 
the other one due to the experimental errors' sources. 

Furthermore, the numerical analysis of 
the oscillation period, as function of the scattering length, suggests a new method for the experimental 
measurement of the scattering length of BEC's atoms.

\acknowledgments

This work is partially supported by Gruppo Nazionale per la Fisica Matematica (GNFM-INdAM).

\end {document}